# Parallel Closed-Loop Connected Vehicle Simulator for Large-Scale Transportation Network Management: Challenges, Issues, and Solution Approaches

Mohammad A. Hoque, *Senior Member, IEEE*, Xiaoyan Hong, *Member, IEEE*, Md Salman Ahmed, *Student Member, IEEE*

*Abstract*— The augmented scale and complexity of urban transportation networks have significantly increased the execution time and resource requirements of vehicular network simulations, exceeding the capabilities of sequential simulators. The need for a parallel and distributed simulation environment is inevitable from a smart city perspective, especially when the entire city-wide information system is expected to be integrated with numerous services and ITS applications. In this paper, we present a conceptual model of an Integrated Distributed Connected Vehicle Simulator (IDCVS) that can emulate real-time traffic in a large metro area by incorporating hardware-in-the-loop simulation together with the closed-loop coupling of SUMO and OMNET++. We also discuss the challenges, issues, and solution approaches for implementing such a parallel closed-loop transportation network simulator by addressing transportation network partitioning problems, synchronization, and scalability issues. One unique feature of the envisioned integrated simulation tool is that it utilizes the vehicle traces collected through multiple roadway sensors—DSRC onboard unit, magnetometer, loop detector, and video detector. Another major feature of the proposed model is the incorporation of hybrid parallelism in both transportation and communication simulation platforms. We identify the challenges and issues involved in IDCVS to incorporate this multi-level parallelism. We also discuss the approaches for integrating hardware-in-the-loop simulation, addressing the steps involved in preprocessing sensor data, filtering, and extrapolating missing data, managing large real-time traffic data, and handling different data formats.

*Keywords*—Connected Vehicle, Parallel Simulation, Network Partitioning, Scalability, Communication Overhead, Dedicated Short Range Communication, hardware-in-the-loop simulation.

## I. Introduction

With the advent of big data and connected vehicle (CV) technologies, the parameters and requirements for simulating metro-scale urban transportation networks with heterogeneous vehicles have evolved substantially. Today's transportation engineers at the Traffic Management Centers (TMCs) feel the necessity of a parallel CV simulation tool that could allow them to visualize the immediate system-wide effect of any change in traffic parameters—signal timing, detour, lane closure—before making the decisions. Unfortunately, current state-of-the-art traffic simulators (VISSIM [1], CORSIM [2], SUMO [3] etc.) are designed and developed primarily for the microscopic simulation of vehicles and pedestrians, evaluation of traffic control algorithms, and visualization of on-road traffic behaviors. Since these traffic simulators were not originally designed to model the multi-layer wireless network protocols (IEEE 802.11p, IEEE 1609.x) required to simulate the vehicle-to-vehicle (V2V) and vehicle-to-infrastructure (V2I) communications, additional work is necessary to port and feed the network simulators with these traffic simulators to simulate a transportation network with connected vehicles. Simulation of a transportation network with CV requires a bi-directional coupling mechanism between a transportation simulator and a communication simulator. This mechanism has led to the concept of the closed-loop CV simulator, which has recently drawn a significant amount of research interests within the community. However, the computational capacity of such a bi-directionally coupled (closed-loop) simulator is significantly limited by the number of CVs equipped with onboard units (OBUs) and the number of roadside units (RSUs) deployed within the metro-wide transportation network, since these DSRC devices transmit millions of basic safety messages (BSMs) packets every minute requiring massive computational resources. Existing sequential closed-loop simulators can barely handle one thousand vehicles simulated in a scenario involving no more than a few intersections. Thus, the incorporation of parallelism in both transportation and communication simulation platforms will enable efficient management of large-scale transportation network and control of traffic parameters involving connected vehicles. In addition, the integration of roadway sensor data through hardware-in-the-loop simulation with the closed-loop software simulator will enable the traffic engineers to make informed decisions by evaluating the system-wide impact of traffic parameters changes in real-time. The hardware-in-the-loop simulation (HILS) module in a parallel computing environment will facilitate system-wide visualization of traffic status from the Transportation Management Center (TMC). A potential use case for our HILS-incorporated parallel CV simulation tool is to provide a realistic prediction of the consequences of traffic change—such as transit bus delays or tentative queue length considering the preemptive detour advisory disseminated through DSRC—enabling a traffic engineer to make the real-time decision when a major corridor needs to undergo closure of lanes due to maintenance. Another important benefit of integrating HILS is to validate the simulation results of V2I applications.

A vast amount of research effort has been recently directed towards the improvement of surface transportation through self-driving autonomous vehicles as well as connected vehicles (CVs) using the 5.9 GHz Dedicated Short Range

Communication (DSRC) technology. Automakers and technology developers like Google, Ford, and General Motors etc. are working to improve the controllability features of autonomous or semi-autonomous vehicles. While self-driving cars can potentially reduce the stress of navigating through congested traffic, CVs can optimize the traffic flow across an entire transportation network through the exchange of information among vehicles and infrastructure. CV applications use information obtained through V2X communications to assist drivers in avoiding congestion, reducing vehicle stops, choosing the best route, and optimizing fuel efficiency. Hence, CV-based emerging Intelligent Transportation Systems (ITS) applications can result in transformative changes to the overall surface transportation system.

To accurately simulate ITS applications on a scenario involving connected vehicles, it is necessary to integrate a full-fledged transportation simulator with a wireless network simulator, resulting in the need for a closed-loop simulator. This kind of closed-loop simulator requires a tight synchronization between two stand-alone simulation modules: a transportation module and a communication module. The transportation module is responsible for the modeling of vehicle mobility applications including traffic routing, car-following, lane-changing, vehicle dynamics, driver behavior modeling, and traffic signal control modeling etc. On the other hand, the communication module accounts for data traffic network modeling including packet routing, end-to-end V2X packet delivery, wireless media access, cross-layer protocols, information security, and authentication mechanisms.

In a CV simulation environment, the two simulation modules (transportation and communication) operate as a real-time feedback control loop with a tight synchronization. These two modules highly influence the operations of one another. For example, vehicle dynamics, mobility, speed, and density affect the communication links between vehicles as well as the data packet routing; hence, they also affect the communication quality, i.e., reliability, throughput, and delay. Conversely, the data communication parameters—for example, the number of packet losses between vehicles and the end-to-end delivery delay—can adversely affect the mobility decisions made by the transportation simulator, particularly when a V2X message carries detour information due to an accident. For a V2X-based safety application, it is important to realize that the slightest delay in communication, even about a fraction of a second, can have serious consequences and may even be fatal. Considering the complexity of the transportation and communication module with the high level of interdependency between them, it is easy to perceive how challenging the simulation of an integrated CV system can be.

## II. RELATED WORK

Researchers have been focusing to develop a complete feedback-loop based transportation simulator for the past decade. More specifically, they are putting their effort to develop a transportation network with a wireless network simulator for simulating V2X-based ITS applications. Many researchers studied sequential simulators, but a comparative modular analysis of different simulator components is still needed for identifying the capabilities and limitations of the simulators. Thus, we focus on several sequential simulators to compare their modular organization and architecture.

Early efforts to simulate vehicular networks were based on fixed mobility trajectories that were fed to the network simulators. Contemporary network simulators are designed to simulate communication protocol and hence, do not have the capability to generate a realistic car-following model. Thus, there was a need for ingestion of fixed mobility traces for the communicating nodes. Several mobility generator frameworks (VANETMOBISIM [7], SUMO [3], MOVE [8], STRAW [9], FREESIM [10], CITYMOB [11], C4R [38]) have been developed to produce the vehicular trajectories that are fed into various network simulators (NS2 [12], NS3 [13], OMNET++ [14], OPNET [15], JIST/SWANS [16], QualNET [17], etc.). However, ingesting static mobility traces into these network simulators could not incorporate the effect of ITS applications on the mobility of the vehicles.

A. *Modular Analysis*

Few sequential simulators such as OVNIS [27], TraNS [18], and iTETRIS [25] use SUMO [3] as their traffic model and NS-3 [13] as their communication model. But their fundamental functionalities are different. For example, node application module makes OVNIS different from the rest two. OVNIS's node application module has the ability to query the traffic model. TraNS utilizes the same set of traffic and communication models for simulating static network (i.e., no alteration of traffic flow) and dynamically generated network (i.e., alteration of traffic flow by abrupt braking and collision avoidance). iTETRIS focuses on simulating more complex traffic scenarios than the previous two. It also the first simulator that supports many standards (WiMAX, UMTS, DVB-H, and ETSI) and external module integration.

Though Veins [20] uses SUMO as its traffic model, it uses OMNET++ [14] as its communication model for evaluating inter-vehicle communication (IVC) protocols on-road mobility. For example, Veins can suggest alternate routes by simulating accidents or crashes using the IVC protocols.

Application-aware SWANS with Highway mobility (ASH) [26] integrates Scalable Wireless Ad hoc Network Simulator (SWANS) [16] as its network model and Intelligent Driver Model (IDM) [61] as its traffic model. ASH supports a two-way communication between the traffic mobility and networking models. It also supports an Inter-Vehicle Geocast (IVG) [62] based broadcasting technique.

The primary goal of VnetIntSim [22] is a movement-based simulation. VnetIntSim uses INTEGRATION [63] as its traffic model and OPNET [15] as its communication model. The simulator can simulate vehicle-to-vehicle and vehicle-to-infrastructure scenarios.

Many simulators use their own traffic and communication models or provide widespread external module integration for supporting different simulations. For example, GrooveSim [19] utilizes their own models to support an on-road driving mode, a virtual traffic network simulation mode, a playback mode, a hybrid simulation mode, and a test scenario generation mode. Automesh [64] supports external modules integration with plug and play capabilities. However, the framework was merely proposed but not implemented. STRAW [9] utilizes their own traffic model for simulating intra-segment and inter-segment vehicle motion.

### B. Application Focused Simulation

Some researchers have been focusing on developing simulation frameworks for simulating intelligent traffic controls for autonomous vehicles. For example, Gelbal et al. present a hardware-in-the-loop (HiL) simulator capable of running lane keeping and adaptive cruise control car following applications [65]. The HiL simulator consists of a dSPACE Scalexio system, a dSPACE Microautobox control unit, DSRC radios, and sensors. The dSPACE Scalexio system runs the Carsim simulator with real-time traffic and sensor data. The dSPACE Microautobox control unit provides the autonomous functionalities. So, the HiL simulator focuses more on the application side of a simulator rather than the internal modular design.

Some researchers also focus on the application side of a connected vehicle simulator. Luigi et al. compare the spacing adopted in equilibrium car-following conditions involving a real driver in the loop simulator using driving simulators and instrumented cars [66].

Many researchers have been trying to integrate simulators in different domains to package those simulators into a single simulation framework. For example, Zhao et al. propose a 3-1 integrated traffic-driving-networking simulator (ITDNS) [67]. The purpose of ITDNS is to simulate cyber transportation system and connected vehicle applications allowing a human driving into the virtual simulation environment. The authors in [68] discuss a framework for simulating a road traffic control policy to reduce the waiting time for emergency vehicles. To do so, they interconnect a multi-agent system development framework (JADE) and an agent-based traffic simulator (SUMO) using TraSMAPI [69]. IsV2C, an integrated road traffic-network-cloud simulator [70], aims to simulate the vehicle-to-cloud services for providing real-time performance for applications such as driving assistance, infotainment, and vehicle maintenance. Another service-oriented simulator, QoS-CITS [71] provides researchers ways to conduct various experiments including parameter tuning for their study. The authors in [72] discuss Similitude, a framework for simulating traffic scenarios, network communications, and ITS mobile applications. They use SimMobility as the traffic simulator, ns-3 as the network simulator, and QEMU as the mobile application (Android) simulator.

Many researchers also enhance existing traffic and network simulators to cover a wide range of traffic scenarios and communication mediums. For example, researchers in [73] focus on simulating network communications in heterogeneous networks keeping in mind to support new communication mediums as well as the legacy mediums. More specifically, they extend ns-3 to support the communication simulation in visible light communications (VLC) medium. Lim et al. enhance SUMO for meeting requirements for the roadway scenario generation for different transportation system (e.g., left-hand side simulation or right-hand side simulation) [74]. Abeywardana et al. extend veins to add support for advertising cognitive radio-based services [75].

### C. Open-Loop Simulators

Lee and Park [18] used the NCTUns communications simulator to examine the effects of communications using VISSIM trajectory data offline with no feedback loop for traffic simulation. GrooveSim [19] simulates inter-vehicular communication and vehicular mobility in a road traffic network using a customized mobility model and the GrooveNet [20] routing protocol. MobiREAL [21] incorporates mobility support on the Georgia Tech Network Simulator (GTNetS [22]). The capabilities of these type of open-loop simulators are limited to studying only unidirectional effects between the two domains. For example, studying the effect of various mobility models on the performance of end-to-end data communication using these simulators could characterize the dependency of the communication module on the transportation module, but it would be impossible to study the impact of data communication on the transportation system by incorporating changes in vehicle route, speed, signal timings, and mobility patterns based on newly received messages. Hence, this approach cannot be used to study bidirectional effects between the two tightly coupled domains.

### D. Closed Loop Simulators

Recently, there has been a significant amount of interests and efforts to design closed-loop CV simulators by coupling transportation and wireless network simulators. These closed-loop simulators can be further divided into two types—1) Fixed pair coupling and 2) Flexible pair coupling.

*1) Fixed Pair Coupling*

Most closed loop CV simulators are fixed pair, which means that the choice of transportation simulator and communication simulator is already decided based on the CV simulator. For example, Traffic and Network Simulation Environment (TraNS [23]) links SUMO and NS-2. Multiple Simulator Interlinking Environment for IVC (MSIE [24]) integrates NS-2, VISSIM traffic simulation, and application simulation (MATLAB) into a simulation environment for vehicular ad hoc networks (VANETs). Veins [4] provides a closed-loop integration using SUMO and OMNeT++ as the traffic and communication simulators respectively. Integrated Wireless and Traffic Platform for Real-Time Road Traffic Management Solutions (iTETRIS [25]) integrates SUMO with NS-3 through IP-based sockets and allows implementation of several ITS applications in various programming languages. Very recently, Songchitruksa et. al. developed a closed-loop CV simulator (CONVAS [5, 42]) by interlinking VISSIM and NS-3. VNetIntSim [6], another recent closed-loop simulator, couples OPNET and INTEGRATION. A comprehensive survey of contemporary CV simulators can be found in one of our earlier papers [43].

*2) Flexible Pair Coupling*

In contrast to existing fixed simulator couplings, the VSimRTI [76, 77, 78] simulation platform allows the flexible integration of traffic, communication, and emission simulators. Therefore, the high modularity of VSimRTI enables the coupling of the most appropriate simulators for a realistic representation of transportation, emissions, communication, driver behavior, and mobility modeling. Based on the requirements of a particular simulation scenario, the most relevant simulators can be chosen. In addition, VSimRTI extends the simulation of electric vehicles. VSimRTI utilizes an ambassador concept to couple arbitrary simulation systems with a remote-control interface. In order to attach an additional simulator, the ambassador interface needs to be implemented. Currently, VSimRTI already includes an interface for traffic simulators SUMO and PHABMACS and the communication simulators ns-3 and OMNeT++.

Below we have summarized all these simulators in the Table 1:

**Table 1: Summary of Integrated CV simulators**

| Integrated Simulator | Traffic Simulator | Network module | Open-loop vs Closed-loop |
|---|---|---|---|
| NCTUns [18] | VISSIM | NCTUnc | Open-loop |
| MOBIREAL [21] | CPE model | GTNeTS | Open-loop |
| GrooveSim [19] | Own model | GrooveNet | Open-loop |
| ASH [26] | IDM/MOBIL, IVG | SWANS | Closed-loop |
| OVNIS [27] | SUMO | NS-3 | Closed-loop |
| Veins [20] | SUMO | OMNET++ | Closed-loop |
| VnetIntSim [22] | INTEGRATION | OPNET | Closed-loop |
| TraNS [18] | SUMO | NS-2 | Closed-loop |
| iTETRIS [25] | SUMO | NS-3 | Closed-loop |
| CONVAS [23] | VISSIM | NS-3 | Closed-loop |
| VSimRTI[76,78] | Flexible | Flexible | Closed-loop |

Unfortunately, none of these closed-loop simulators described above integrate with hardware-in-the-loop simulation technique. These tools also lack in providing support for simulating large-scale transportation scenario using parallel and distributed computing. Another major limitation is that there is no mechanism available for collecting roadway sensor data from individual intersections and feeding them to the simulation environment to facilitate real-time traffic decision support at the TMCs.

### III. PARALLEL SIMULATION OF CONNECTED VEHICLE APPLICATIONS

#### A. *Need for Parallelism*

The augmented scale and complexity of urban transportation networks have significantly increased the execution time and resource requirements of vehicular network simulations, exceeding the capabilities of conventional simulators. The need for a parallel and distributed CV simulation environment is inevitable from a smart city perspective where the entire city-wide information system will be integrated with numerous services and ITS applications, particularly when the metro-wide multimodal transportation systems get connected to the smart city infrastructure through DSRC.

Unfortunately, contemporary simulation tools do not provide any mechanism for parallel or distributed simulation of CV applications for large-scale transportation networks. Due to the complexity of implementation, earlier attempts [46, 47] of parallelizing vehicular network simulation were limited to only open-loop simulators, which do not facilitate real-time feedback control between the transportation and communication parameters. To address the need for ITS practitioners and researchers, we develop a framework for parallelizing closed-loop CV simulation, with the option to incorporate real-time roadway sensor data through hardware-in-the-loop simulation.

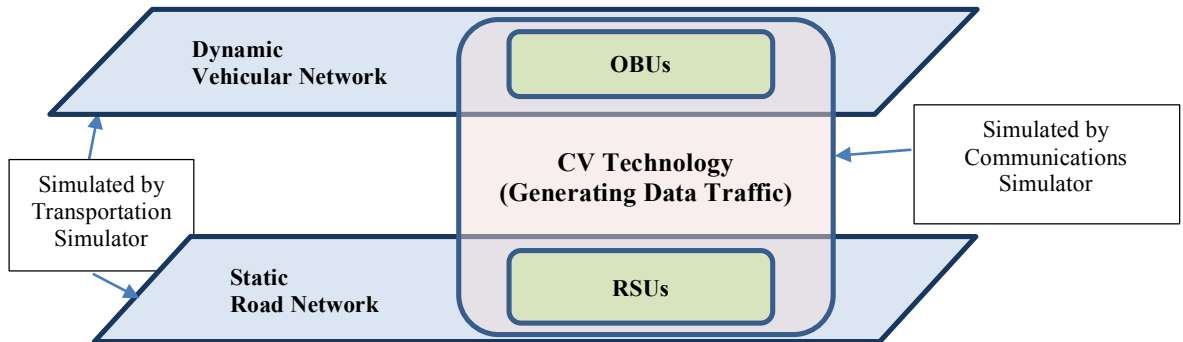

Figure 1: Bi-layer complex transportation network

#### B. *Implementation Challenges and Issues*

In this section, we identify some of the challenges and issues associated with implementing a parallel closed loop simulator for the large-scale transportation network management. Later we provide insights into the solution approaches that can address these problems.

*1) Partitioning of Bi-layer Complex Transportation Network*

The fundamental research problem involved in this parallel simulator design is to determine a near-optimal partitioning heuristic using a bi-layer network model—a static road network overlaid with a dynamic vehicular network—connected by the CV technology which spans across both the networks (Figure 1). Earlier research mainly focused on partitioning static road networks for distributed simulation without considering the data traffic generated by DSRC communications. The bi-layer model will address partitioning issues in both the dynamic vehicular network involving CVs (OBUs) as well as the static infrastructure (RSUs) and the interactions between the two levels. The solution approaches in section IV will provide guidelines to incorporate real-world roadway traffic parameters with the data traffic parameters within the partitioning heuristic for connected vehicle environments.

The biggest challenge in partitioning vehicular networks is that the partitions cannot be fully separated. In fact, due to the communication and high mobility, partitions have a high level of interdependency and interactivity (i.e., a message or a vehicle moves from one partition to another) that necessitate communications between partitions to achieve consistency and accuracy. Inefficient partitioning of such networks can produce high communication volume between partitions and high processing overhead within a partition, consequently resulting in low simulation speeds. So, it is necessary to create partitions in such a way that reduces the interactivity and interdependence between them. Another proven NP-hard problem is the load-balancing problem. Due to the interdependency between events in different partitions, the simulation must be synchronized between the partitions; i.e., low-load (high-speed) partitions must wait for high-load (low-speed) ones to finish. This means that the maximum overall simulation speed is limited to the minimum speed among all the partitions.

*2) Reducing Inter-Simulator Communication Overhead*

A major problem for parallelizing a closed-loop CV simulator is that it not only requires decomposing the two standalone simulators (the transportation and communication simulators) and synchronizing the components within each simulator, but it also requires a tight synchronization between the two simulators. DSRC technology requires that the vehicles broadcast their current locations every 0.1 seconds, meaning that the two simulators must synchronize ten times per second. This synchronization process adds extra overhead if the two simulators are running on separate computing nodes in a distributed computing environment requiring them to communicate over the Message Passing Interface (MPI). With a shared memory interface between the two simulators running in the same partition, this Inter-Simulator Communication overhead is expected to be reduced. However, using shared memory also creates a race condition between multiple processes running on the same computing node. Hence, there is always a trade-off between contention (shared memory) and latency (distributed memory), which is a major research problem. In addition, in a CV environment, the closed-loop interactions between communication and transportation systems must be executed in real-time to accurately model the impact of one system on its counterpart. For instance, the real-time interactions between SUMO and OMNET++ should facilitate dynamic speed control for the vehicles in the vicinity of traffic signals, where vehicles and signal controllers can exchange information to compute the optimal signal timing and vehicle trajectory.

*3) The existence of heterogeneous vehicles*

Another challenging aspect of simulating transportation network involving CVs is due to the slow market penetration rate of connected vehicles, which implies that during the transition period there will always be two types of vehicles on the road—one that is connected through DSRC (CV) and the other that is not connected (non-CV). It is expected that CV technologies will penetrate the market slowly over the next few years. Hence, until the time comes when all the cars on the road are equipped with factory-built or after-market DSRC devices, there will always be two types of vehicles on the road: one that has DSRC onboard unit (OBU) and the other that does not have OBU. CVs broadcast their actual GPS positions and speed every 0.1 seconds through the basic safety messages (BSMs). So, CVs can be easily identified through the BSM packets. However, detection of non-CV vehicles may not be easy, although it is possible to detect the non-CVs and reconstruct their mobility traces using several techniques. First, we can equip optical sensors to CVs so that the CVs can collect surrounding vehicles' information using the optical sensors. A CV can detect duplicate vehicles by communicating with other CVs. Second, we can reconstruct the mobility traces using some established techniques discussed in [39], [40], and [41]. Third, the data from the loop detector and traffic cameras can further refine non-CV detection and mobility trace reconstruction. At present, there is no closed loop simulator that supports the simulation of CV applications with heterogeneous types of vehicles.

Using the state-of-the-art closed-loop simulator with the support for hardware-in-the-loop simulation, vehicles with attached OBUs will be able to participate in the network-wide communication, while the vehicles without OBUs will not be detected by the simulator. Incorporating non-CVs in the hardware-in-the-loop simulation is quite challenging because the closed-loop CV simulator needs to be fed from several sources of sensor data such as CV traces through BSM messages and non-CV traces from roadway sensors (loop detectors and video detectors). From the input of these sources, the simulator needs to generate realistic mobility traces for the non-CVs, in addition to mapping the actual positions of the CVs where the simulator should graphically represent the CVs and non-CVs differently to distinguish between the actual position and speed vs. projected position and speed.

*4) Synchronization problem*

Simulation of data traffic is computationally more resource intensive than the simulation of vehicular traffic [6]. This makes the closed-loop simulation of CVs challenging because the imbalance of computational resource requirement causes synchronization problems between the transportation simulator and communication simulator. The synchronization problems happen due to the huge amount of DSRC basic safety messages (BSM) disseminated from each vehicle in every tenth of a second, where each BSM message needs to go through several layers of encapsulation and de-capsulation steps within the wireless network's protocol stack at both ends. Some of the services in the data communication protocols, e.g. error detection, routing, and connection establishment, are computationally more expensive compared to the services from the vehicular traffic simulator that do not require passing through multiple layers of protocols. In fact, simulation of vehicular traffic only involves trace generation using microscopic mobility models. Hence, the data traffic simulator primarily causes the bottleneck. Typically, the data traffic is simulated using network simulators such as OPNET, OMNET++, Qualnet, NS-2, or NS-3. One experiment in [29] demonstrated that the simulation of a 200-node network for only one minute generated more than 4,600,000 events and required 16 minutes of CPU time. The increasing complexity of the protocol stacks in end devices further aggravates this problem and has spurred efforts to develop parallel network simulators.

### 5) Scalability of Parallel Simulation

The scalability of parallel systems depends on the ratio of time spent in computation vs. communication. For any parallel system, the fraction of time spent in inter-process communication increases with the number of processors while the fraction of time spent in actual computation decreases. Initially, the computational time is greater than the communication time for a lower number of processors. The computational time decreases with respect to the communication time with the increase of the number of processors. At some point, for a specific number of processors ($p$), the communication time starts dominating over computation time. This value of $p$ essentially determines how better the system is scalable. The higher the value of $p$, the better the scalability. Therefore, scalability is one of the most important problems in any parallel system, especially when it involves both distributed and shared memory architecture. Hence the architecture of such hybrid parallel system needs to be designed in such a way that reduces the inter-process communication (IPC) overhead and increases the scalability. It is noteworthy to mention that this IPC could take place between the transportation simulator and network simulator using shared memory (using OpenMP) or between the instances of the same simulator running different partitions on distributed cluster nodes (through MPI). Without achieving a certain level of scalability, the system will not be able to simulate a city-wide scenario with several hundred thousand vehicles and millions of BSM messages in every minute. Ahmed et. al. studied the scalability issues in terms of memory usage and execution time using VNetIntSim [33, 34] proving that the number of wireless nodes (vehicles) and the data traffic rate per vehicle are the primary reasons behind the scalability issue.

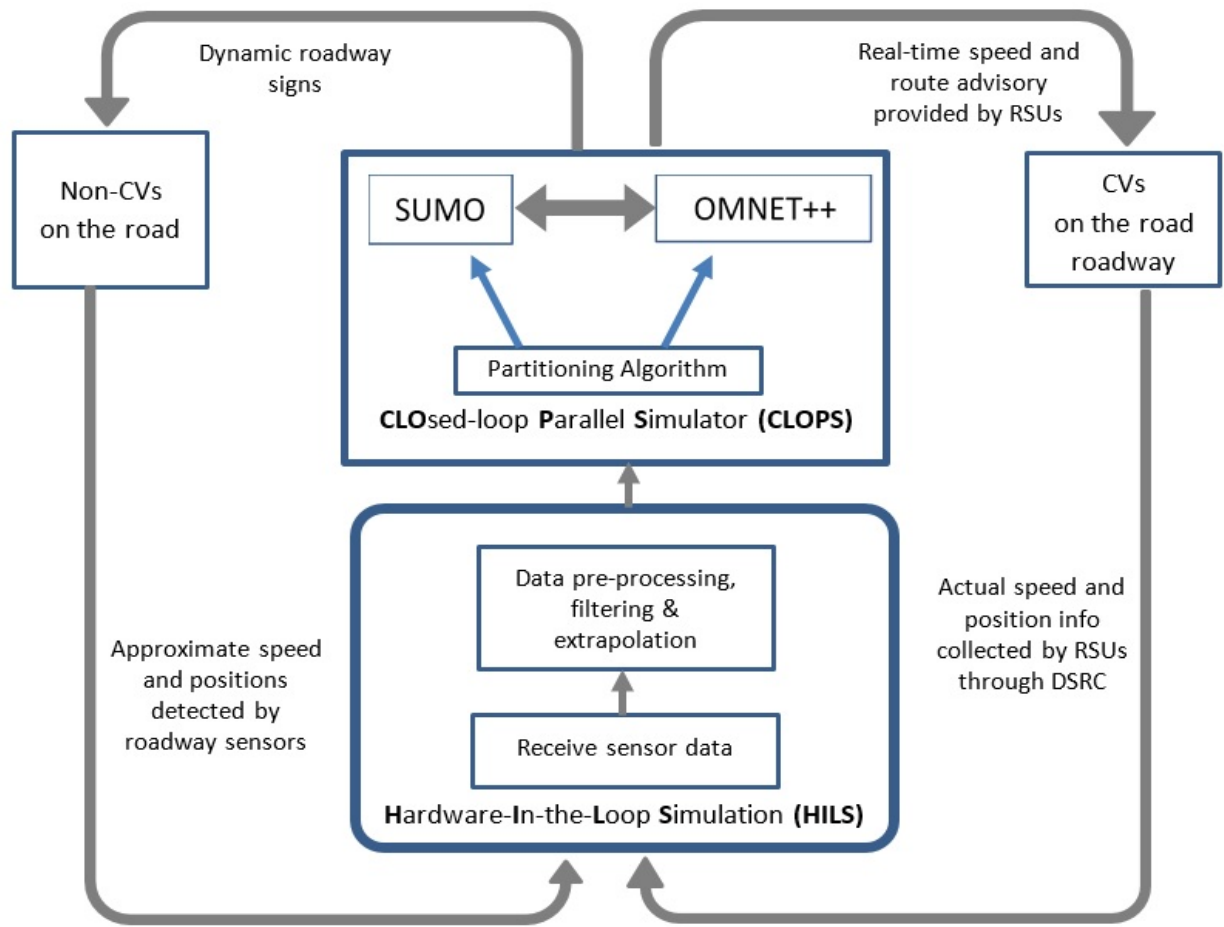

Figure 2: Conceptual Framework of Integrated Distributed CV Simulator (IDCVS)

## IV. Conceptual Model

In this section, we present a conceptual framework of an **I**ntegrated **D**istributed **C**onnected **V**ehicle **S**imulator **(IDCVS)** and in section V we discuss the technical approaches for implementing such a robust simulation tool. Figure 2 below shows our hypothetical model of IDCVS system that includes hardware-in-the-loop simulation techniques for both CVs and non-CVs. IDCVS will incorporate two basic modules—a **CLO**sed-loop **P**arallel **S**imulator (**CLOPS**) and a **H**ardware-**I**n-the-**L**oop **S**imulation (**HILS**) module.

### A. *Hardware-In-the-Loop Simulation (HILS)*

HILS will have an interface to receive the sensor data from both CVs and non-CVs through multiple sources. For non-CVs, the approximate location and speed can be detected through video detectors and inductive loop detectors, and this information will be passed as input to the HILS module. We can use the video detection and the loop detector software that can supply the sensor data to the HILS receiver component. On the other hand, the CV's can be detected easily through the BSM messages received by the RSUs. Once the sensor data is received, additional data-preprocessing, filtering, and extrapolation will be needed before the data can be used by CLOPS. This will require developing filtering algorithms for loop-detection and video-detection data to isolate the CV traces from the non-CV traces.

### B. *CLOsed-loop Parallel Simulator (CLOPS)*

A CLOsed-loop Parallel Simulator (CLOPS) can be developed by interlinking OMNET++ with SUMO where both simulators are open source. An efficient partitioning heuristic will decompose the complex transportation network into two separate sets of partitions—where each set of the partition will be sent to the individual simulator (SUMO and OMNET++). It might appear that CLOPS could be developed as a parallel and distributed framework on top of Veins since Veins also utilizes a coupling between SUMO and OMNET++. However, since Veins does not support heterogeneous vehicles, it is not possible to extend Veins for the simulation scenarios involving both CVs and non-CVs. In addition, CLOPS may have the capability to vary the ratio of CVs to the non-CVs as per the market penetration rate. This requires a non-uniform partitioning between SUMO and OMNET++.

### C. *Modes of Operation for IDCVS*

One important feature of this conceptual IDCVS system is that it will have the option to simulate in two different modes—closed-loop simulation (CLSim) mode and HILS-mode. The CLSim mode will simulate without sensor data, in this case the entire simulation will be run within CLOPS. To incorporate both DSRC-equipped and non-equipped vehicles on a CLSim scenario, we can randomly distribute the CV vehicles within the road network based on a user-specified technology penetration rate. On the other hand, the HILS-mode will enable simulation based on real-time sensor data.

## V. Implementation Approaches

In this section, we discuss the possible technical approaches to address the challenges pertaining to implementation of the integrated simulator.

### A. *Developing Network Partitioning Heuristic*

A crucial challenge for the partitioning problem described in section III is that, due to the imbalance of computational resource requirements between transportation simulator (SUMO) and network simulator (OMNET++), a single partitioning scheme may not work for both of the simulators. Apart from that, the number of vehicles will also vary among the two simulators where SUMO needs to simulate the traces for all vehicles (both CVs and non-CVs) whereas OMNET++ only simulates data traffic generated from the CVs. If a single partitioning heuristic is used, the synchronization problem will be further aggravated. Hence, it is necessary to have two separate partitioning schemes for SUMO and OMNET++.

A Boolean matrix based vehicular network partitioning algorithm has been developed by Hoque et. al [44, 45] which is expected to be incorporated into our distributed simulation architecture involving hybrid parallelization model. In our recent work related to road-network partitioning [30, 31], we have identified the following issues and parameters that play vital roles in designing an efficient partitioning heuristic:

*1) System boundary nodes of each partition:* The total number of inter-process communication or messaging depends on the number of system boundary nodes of each partition.

*2) The number of partitions:* Almost every graph partitioning algorithm is based on a pre-specified number of partitions, which may not always generate the optimal solution in practice. Instead of specifying an exact number of partitions, an upper bound and lower bound can be provided as input to the algorithm to determine the best partitioning solution within the specified range.

*3) Intersection cut:* If an intersection is considered as a boundary node for a partition, then a significant amount of vehicle mobility data must be communicated between the partitions. In this context, an important factor—whether to prioritize signalized intersection over un-signalized intersection as a candidate for boundary node—remains open for further research, which should be investigated.

*4) Link/Edge cut:* When a link or edge is selected to be cut then the traffic volume along the cut link is directly proportional to the amount of information exchanged between the two partitions along the link. In this case, a good strategy would be to cut the links with minimum traffic to reduce the communication overhead between partitions.

## B. *Partitioning Approach for SUMO*

To create the network graph, the OSM file of the experimental city can be downloaded from the www.openstreet.org website. To avoid unnecessary complexities, residential street, service path, footway, cycleway, motorway, and unclassified roads can be excluded from the graph. Table 2 shows some suggested parameters that can be incorporated to generate the weighted graph matrix. It could be easily possible to extend an existing graph partitioning software like METIS [37] for generating the partitions of the transportation network. Figure 3 shows our preliminary transportation network partitioning using the above-mentioned parameters with the help of METIS. METIS is a very stable partitioning package implementing the popular Kernighan-Lin heuristic. METIS divides a graph into three phases: coarsening, partitioning, and uncoarsening. In coarsening phase, the heavy edge matching scheme can be used, whereas, in the uncoarsening phase, the Kernighan-Lin graph refinement algorithm can be used. The coarsest graph can be bisected using graph growing followed by boundary Kernighan-Lin algorithm with graph partitioning using recursive bisection technique. The input for METIS can be provided using the generated graph matrix and weight parameters.

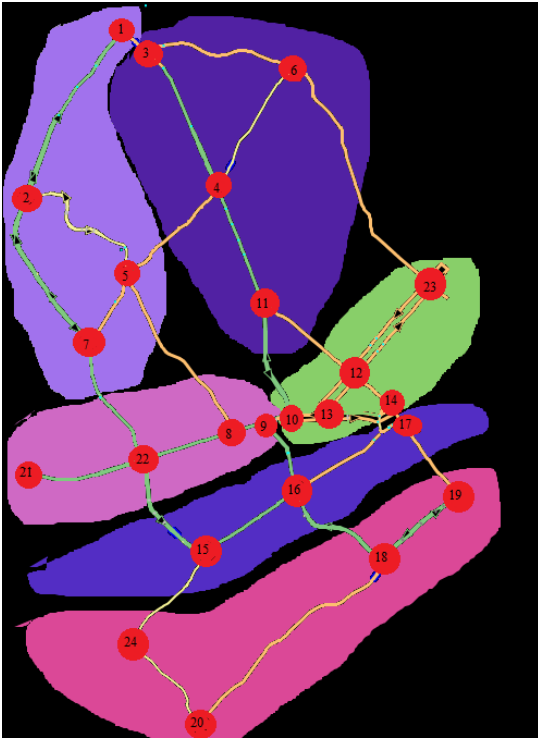

Figure 3: Partitions of the transportation corridor of Johnson City, Tennessee, United States

Table 2: List of parameters considered for partitioning heuristics

| Parameter | Technique |
|---|---|
| *Node weight* | All signalized intersections in the OSM data will be identified. These types of intersections or nodes will be assigned a higher weight. Un-signalized intersections will be assigned the sum of the number of incoming and outgoing lanes as the weight. |
| *Link length* | The length between two nodes will be calculated using the Haversine formula: $$d = 2r \sin^{-1}\left(\sqrt{\sin^2\left(\frac{\varphi_2 - \varphi_1}{2}\right) + \cos(\varphi_1)\cos(\varphi_2)\sin^2\left(\frac{\lambda_2 - \lambda_1}{2}\right)}\right)$$ where d=Distance between two points/nodes, r=Radius of Earth (6367 km) $\varphi_1, \varphi_2$=Latitude of point 1 and 2 and $\lambda_1, \lambda_2$= Longitude of point 1 and 2 |
| *Number of lanes* | The number of lanes of a road segment or a link will be extracted from the OSM data. |
| *Link density* | The density of a road segment or link will be extracted from the Google Map Application's newly introduced traffic layer [36]. The density is expressed in three categories: low, medium, and high. |
| *Link priority* | The road segment will be assigned a priority index based on the weighted summation of link length, the number of lanes, and link density. |

## C. *Partitioning Approach for OMNET++*

Some of the key factors concerning partitions for OMNET++ in the context of closed-loop parallel simulation for reduced interactivity and interdependence include vehicle mobility, communication events and external stimuli from the simulated transportation applications. These factors directly influence the previously mentioned challenges pertaining to network partitioning. The application stimuli are the drive for CV communications, which can be sporadic or proactive. The transportation network information such as the road network (road links, road nodes), vehicle density on each link, and vehicle speeds and distribution determine the vehicular mobility. This information can be further utilized to quantify the number of communication events. The approach to optimize partitions is to consider the number of discrete events in the communication network as the basis for drawing the boundary between the connected components. For example, one way to incorporate this approach is to employ the vehicle density and the length of each link as link weights in partitioning techniques (such as the minimum cut or minimum k-cut algorithms) to partition the network and minimize the interactivity between different portions. The lower the density and the longer the length of a link, the higher the possibility that the link is a cut link in the network. The rationale is that the density and length represent the continuity of the communication route on this link. Therefore, the lower this ratio (density/length), the less communication between the ends of the link. In addition, the partitions need to be adaptive to the dynamics of the application stimuli and the mobility. To address this issue, we can consider the simulation granularity and duration of the current partition time. The goal is to develop an intelligent algorithm to schedule the partitioning job.

D. *Design of Closed-loop Parallel Simulator (CLOPS)*

The closed-loop parallel simulator (CLOPS) integrates SUMO and OMNET++ as two standalone simulators. OMNET++ has the flexibility to dynamically create and delete nodes; this capability is necessary for a parallel simulation environment since the wireless vehicular nodes will be distributed in multiple network partitions based on geographic location. In addition, OMNET++ provides support for both distributed and shared memory computing which is needed for this project. The PHY and MAC layers of DSRC (IEEE 802.11p and IEEE 1609.4) have already been implemented in the OMNET++ platform by the open-source research community, which can be utilized in our research. This is a big advantage compared with OPNET since OPNET does not currently include the DSRC protocol stack.

CLOPS will incorporate hybrid parallelization schemes for both the traffic simulator and network simulator. This hybrid parallelization schemes will allow the integrated platform to run in parallel on clusters of computers within a supercomputing facility. The hybrid inter-process communication will be incorporated using MPI and OpenMP. In our preliminary work in [32], we investigated the impact of communication overhead on the overall simulation time and implemented a new partial-result accumulation pattern for reducing inter-process communication overhead. Figure 4 illustrates the architecture of the envisioned parallel system that incorporates hybrid parallelism. Both transportation and communication simulators will have master controllers (the Transportation Simulation Controller (TSC) and Network Simulation Controller (NSC)) that will coordinate the computational load distribution among the parallel sub-processes. Each of these sub-processes is supposed to simulate a portion of the transportation network defined by the network partitioning scheme. The controller will communicate with the sub-processes using MPI, while a transportation simulator sub-process corresponding to a specific partition communicates with its network simulator counterpart using OpenMP.

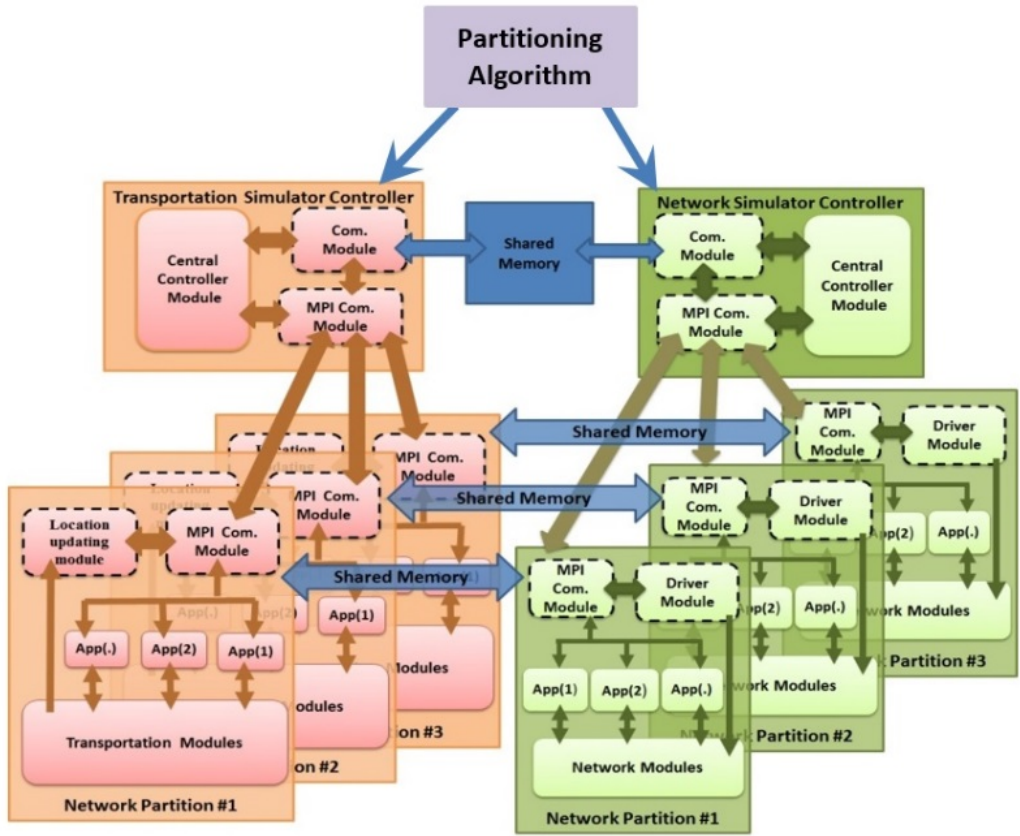

Figure 4. Envisioned distributed simulation architecture incorporating hybrid parallelism

It would be beneficial to utilize two levels of parallelization: network and event levels. At the network level, the overall network can be divided into multiple partitions for both SUMO and OMNET++, each of which will run on a different machine. The TSC and NSC are responsible for managing the loads and synchronizing the partitions within the transportation and communication domains, respectively. At the event level, events can run in parallel within a pre-calculated look-ahead interval. The calculation of the optimum look-ahead interval is crucial in the event-level parallelization. In fact, the look-ahead interval involves a tradeoff between the simulation speed and output accuracy. In the event-level parallelization, utilizing parameters such as node locations and the number of hops between two nodes can increase the scalability of the parallel simulation. For instance, nodes that are spatially separated by long distances can run events in parallel within longer look-ahead intervals without affecting the output.

The communication between the TSC and NSC can be achieved by using shared memory. The vehicles' locations will be calculated and sent to the NSC periodically through the shared memory, and any required application information between the TSC and NSC will be exchanged through the shared memory. Compared to TCP/IP message passing, shared memory has the advantages of reliability and the highest possible speed of information exchange. In contrast, the message size in TCP/IP message passing is limited; thus, in the case of large network size, a large number of messages are needed for each location update. Consequently, TCP/IP message passing may create a communication bottleneck, resulting in the degradation of simulation speed.

E. *Incorporating Hardware-In-the-Loop Simulation (HILS)*

The benefits of incorporating hardware-in-the-loop simulation in the parallel simulation framework are twofold: (a) real-time visualization of system-wide traffic and (b) fidelity testing of V2I applications. They are described as follows.

*a) Visualization:*

One of the purposes of integrating hardware-in-the-simulation (HILS) is to provide real-time traffic information obtained through various roadway sensors to the ITS practitioners monitoring roadway conditions from TMC (traffic management center). The HILS module aggregates roadway sensor data to facilitate visualization of system-wide traffic. A potential use case for our HILS-incorporated parallel CV simulation tool is to provide a realistic prediction of the consequences of traffic change—such as transit bus delays or tentative queue length considering the preemptive detour advisory disseminated through DSRC—enabling a TMC official to make an informed decision when a major corridor needs to undergo closure of lanes due to maintenance.

*b) Fidelity Testing:*

Another important benefit of integrating HILS is to validate the simulation results of V2I applications. For example, eco intersection approach applications [48-60, 79,80] using signal phase and timing (SPaT) information could be first simulated in the CLOsed-loop Parallel Simulator (CLOPS) and then validated using actual roadway sensor data through HILS.

To capture the movement of the non-CVs, several types of detectors can be used such as the magnetometer, inductive loop detection (ILD), video detection, etc. Loop detection is also capable of counting traffic. But it is not 100% reliable for actual traffic counts because the loops in the adjacent through lanes are often tied together for one output for the movement. To solve this problem, the latest video detection technology capable of counting actual traffic can act as a complement for the loop detector. Figures 5(a) and 5(b) shows how the two software detect vehicles at the intersection through software.

Since the target is to simulate both CVs and non-CVs, it is necessary to feed the vehicles' information to the traffic and communication simulators. The RSU can automatically detect the CVs from the BSM packets, but the loop detection and video detection techniques are necessary for detecting the non-CVs. Once the RSU gets the data from all the sources (e.g. BSM packets, inductive loop, video, and magnetometer), a filtering algorithm separates the non-CVs from the CVs using the BSM packets. There is a re-identification issue regarding the use of loop detector, video detection, and magnetometer. In order to eliminate the re-identification problem, we will consolidate the roadway sensor data from three sources along with CV data to accurately identify both CVs and non-CVs. However, detection of the non-CV is not sufficient for the hardware in loop simulation. The mobility trace of a non-CV between two intersections is needed. A car-following model between one/two CVs and a non-CV can be used to extrapolate the missing trace of a non-CV vehicle. For example, the missing mobility trace of a non-CV vehicle can be extrapolated using two CVs' mobility traces where one CV precedes the non-CV and one CV that follows the non-CV. More generally, we can utilize the existing stochastic or probabilistic traffic flow and mobility trace reconstruction techniques [39 - 41] to reconstruct the mobility traces of the non-CV vehicles considering the CV vehicles as the probe vehicles (details discussed in the second challenge of HILS). Other sensors in an intersection can aid to refine the reconstruction. Figure 5(c) shows the flow of sensor data for hardware-in-the-loop simulation.

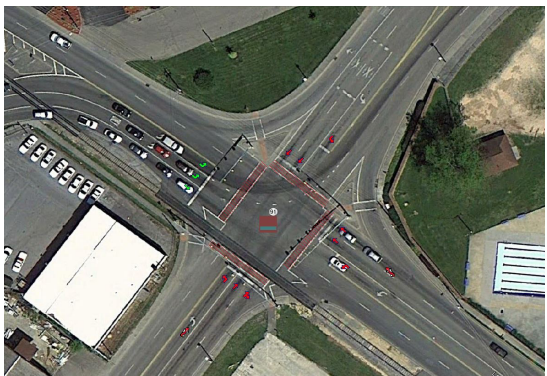

(a) Loop detection software

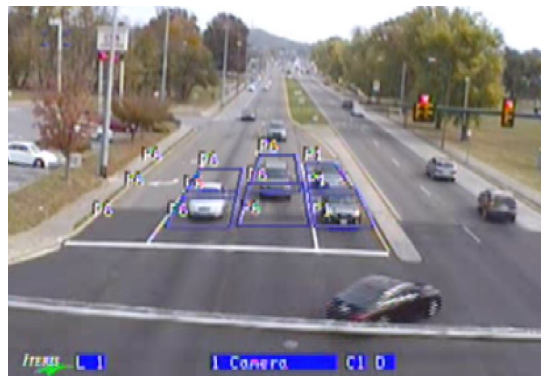

(b) Video Detection software

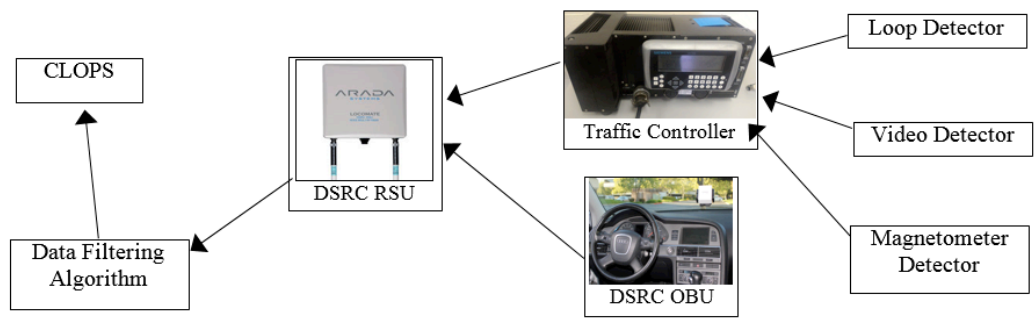

(c) The flow of data between sensors
Figure 5. Integrating hardware-in-the-loop simulation techniques

Some of the challenges associated with integrating hardware-in-the-loop simulation are described below:

*1) Isolating CV traces from loop-detection and video detection data*

The hardware-in-the-loop simulation (HILS) technique can capture roadway sensor data from four different sources—DSRC broadcast messages, inductive loops, video detectors, and wireless magnetometers. Unfortunately, the roadway sensors at intersections cannot differentiate between a CV and non-CV. So, a reliable filtering mechanism is needed to identify the CVs among all the traffic by filtering out the CV data from other two sensors' data based on the GPS position and loop detection timings.

*1) Missing traces between two intersections*

It is very challenging to emulate non-CVs based on sensor data because of the missing traces between two intersections since they can only be detected at the intersections [81-83]. Also, SUMO generated mobility traces between two intersections are the only sources to fill up the missing traces. However, this approach may give some margin of error since some vehicles may arrive at their destination before reaching the next intersection while some other vehicles may start from a mid-point between the two intersections. Since the goal is to approximate the expected traffic between two intersections at a given time, some established statistical models are necessary to validate the simulation results between two intersections. For example, the modal activity-based probabilistic model [39] can be utilized to reconstruct missing non-CV traces using the CV traces. The modal activity sequence in the probabilistic model [39] utilizes the repetitive stop-and-go behaviors of vehicles between two intersections which in fact result in a particular pattern of the activities such as idling, acceleration, cruising, and deceleration. The model generates several vehicle dynamic states using the pattern, ranks the dynamic states using probability distribution functions, and reconstructs the second by second trajectories from the dynamic states. It might be beneficial to incorporate the traffic flow reconstruction methods described in [40] and [41] where each method discusses the traffic flow reconstruction technique using the traffic data such as speeds, positions, and timestamps collected by probe vehicles. In our model, we plan to reconstruct the traces of non-CV vehicles considering the CV vehicles as the probe vehicles. The trajectories of CVs will be reflected in the simulation with almost 100% accuracy given the precision level of GPS sensors integrated with DSRC OBUs. On the other hand, the trajectories of the non-CVs can be deterministically traced near the intersections with the aid of roadway sensors. This means that the non-CVs (real) will be transformed into virtual vehicles between two intersections, while CVs will continue to provide actual mobility traces even within the road segment between two intersections. Between two intersections, the traces of non-CVs can be reconstructed up to a certain level of accuracy by utilizing relative distance information obtained through the sensors equipped in CVs. Even if there is no CV available within the vicinity of a virtual non-CV, still the recent microscopic car-following models can generate a 'close-to-real' trajectory between two intersections in SUMO.

*2) Inaccurate traffic count by loop detectors*

Loop detection can detect traffic but is less reliable for actual traffic counts because the loops in the adjacent through lanes are often tied together for one output for the movement. Also, due to the length of the loop (40 to 50 feet) at the stop bar, multiple vehicles may fall over the same loop or the loops tied together at the same time which reduces vehicle count accuracy.

The current reported accuracy from trace reconstruction model is about 80% [40, 41]. We believe that once we consolidate the sensor data from three different sources and cross match with CV data, the accuracy will be increased. However, some complex vehicle identification algorithm has to be developed to accomplish this goal. The improvement of accuracy level from roadway sensor data is left as an open research issue for the community.

*3) Different data formats*

Typically, data loggers' records include events at an intersection, including a light turning green, a light turning yellow, a vehicle detector turning on, a vehicle detector turning off, and pedestrian walk phase active. While CV data follows DSRC beacon format, loop detector, and video detector inputs are again in a different format. Thus, different pre-processing algorithms are needed.

## CONCLUSION

In this paper, we have discussed a conceptual model that can simulate system-wide changes in traffic parameters on roadways involving both connected vehicles and regular vehicles. We have identified the major challenges and issues for implementing the hardware-in-the-loop simulation and incorporating hybrid parallelism in the closed-loop simulation. We have also discussed the solution approaches for the challenges and issues involved in implementing the conceptual model. However, only a few solutions have actually been implemented. We have discussed possible technical approaches to address the challenges and implementation issues. Our ongoing efforts are directed towards implementation of this model and evaluation of the scalability for emulating metro-wide transportation network. It is noteworthy to mention that a closed-loop integrated parallel simulator may still lack some realistic features within individual simulator modules (e.g. SUMO and OMNET++). However, given that both of the standalone simulators (SUMO and OMNET++) are open-source in nature, it is possible to add the missing features or refine the parameters within each of the simulators (SUMO and OMNET++) even after the implementation of the closed-loop simulator. Our focus in this paper is to point out the challenges of designing the interface between two simulators in the parallel platform, considering the traffic simulator and network simulator as a "black box".

Pre-print (Authors' Version)Medium-Sized Transportation and Communication Networks", Smart Cities, Green Technologies, and Intelligent Transport Systems, Springer Lecture notes in Communications in Computer and Information Science, ISBN: 9783319277530
34. Ahmed Elbary, Hesham Rakha, Mustafa ElNainay, Mohammad A Hoque, "VNetIntSim: An Integrated Simulation Platform to Model Transportation and Communication Networks," International Conference on Vehicle Technology and Intelligent Transport Systems 2015.
35. Do Not Pass Warning Application. Available at: http://www.its.dot.gov/infographs/DoNotPass.htm (*Accessed Nov 13, 2016*)
36. https://productforums.google.com/forum/#!topic/maps/byNxu_lT0do
37. Karypis, George, and Vipin Kumar. "METIS--unstructured graph partitioning and sparse matrix ordering system, version 2.0." (1995).
38. M. Fogue, P. Garrido, F. J. Martinez, J. C. Cano, C. T. Calafate, and P. Manzoni. "A realistic simulation framework for vehicular networks." Proceedings of the 5th International ICST Conference on Simulation Tools and Techniques (SIMUTools), Desenzano, Italy, 19-23 march 2012.
39. Hao, Peng, et al. "Modal Activity-Based Stochastic Model for Estimating Vehicle Trajectories from Sparse Mobile Sensor Data." IEEE Transactions on Intelligent Transportation Systems 18.3 (2017): 701-711.
40. Ramezani, Mohsen, and Nikolas Geroliminis. "On the estimation of arterial route travel time distribution with Markov chains." Transportation Research Part B: Methodological 46.10 (2012): 1576-1590.
41. Zheng, Fangfang, and Henk Van Zuylen. "Urban link travel time estimation based on sparse probe vehicle data." Transportation Research Part C: Emerging Technologies 31 (2013): 145-157.
42. https://tti.tamu.edu/2014/06/01/testing-connected-transportation-innovations-starts-with-first-creating-the-test-itself/
43. M. S. Ahmed, M. A. Hoque and P. Pfeiffer, "Comparative study of connected vehicle simulators," SoutheastCon 2016, Norfolk, VA, 2016, pp. 1-7.
44. Hoque, Mohammad A., Xiaoyan Hong, and Brandon Dixon. "Efficient multi-hop connectivity analysis in urban vehicular networks." Vehicular Communications 1.2 (2014): 78-90.
45. Hoque, Mohammad Asadul, Xiaoyan Hong, and Brandon Dixon. "Analysis of mobility patterns for urban taxi cabs." Computing, Networking and Communications (ICNC), 2012 International Conference on. IEEE, 2012.
46. Lee, Der-Horng, and P. Chandrasekar. "A framework for parallel traffic simulation using multiple instancing of a simulation program." ITS Journal 7, no. 3-4 (2002): 279-294.
47. Luciano Bononi, Marco Di Felice, Marco Bertini, and Emidio Croci. 2006. Parallel and distributed simulation of wireless vehicular ad hoc networks. In Proceedings of the 9th ACM international symposium on Modeling analysis and simulation of wireless and mobile systems (MSWiM '06). ACM, New York, NY, USA, 28-35.
48. Applications for the Environment: Real-Time Information Synthesis (AERIS) Program: AERIS Research Program Overview. URL: https://www.its.dot.gov/research_archives/aeris/index.htm
49. Xia, Haitao, Kanok Boriboonsomsin, Friedrich Schweizer, Andreas Winckler, Kun Zhou, Wei-Bin Zhang, and Matthew Barth. "Field operational testing of eco-approach technology at a fixed-time signalized intersection." In Intelligent Transportation Systems (ITSC), 2012 15th International IEEE Conference on, pp. 188-193. IEEE, 2012.
50. Xia, Haitao. Eco-approach and departure techniques for connected vehicles at signalized traffic intersections. University of California, Riverside, 2014.
51. Hao, Peng, Guoyuan Wu, Kanok Boriboonsomsin, and Matthew J. Barth. "Developing a framework of eco-approach and departure application for actuated signal control." In Intelligent Vehicles Symposium (IV), 2015 IEEE, pp. 796-801. IEEE, 2015.
52. Li, Weixia, Guoyuan Wu, Matthew J. Barth, and Yi Zhang. "Safety, mobility and environmental sustainability of Eco-Approach and Departure application at signalized intersections: A simulation study." In Intelligent Vehicles Symposium (IV), 2016 IEEE, pp. 1109-1114. IEEE, 2016.
53. Zweck, Michael, and Michael Schuch. "Traffic light assistant: Applying cooperative ITS in European cities and vehicles." In Connected Vehicles and Expo (ICCVE), 2013 International Conference on, pp. 509-513. IEEE, 2013.
54. Xia, Haitao, Guoyuan Wu, Kanok Boriboonsomsin, and Matthew J. Barth. "Development and evaluation of an enhanced eco-approach traffic signal application for connected vehicles." In Intelligent Transportation Systems-(ITSC), 2013 16th International IEEE Conference on, pp. 296-301. IEEE, 2013.
55. Cho, Cheng-Hsuan, Han Su, Yi-Hong Chu, Wen-Yao Chang, and Frank Chee-Da Tsai. "Smart moving: A SPaT-based advanced driving-assistance system." In Network Operations and Management Symposium (APNOMS), 2012 14th Asia-Pacific, pp. 1-7. IEEE, 2012.
56. Zhao, Yiran, Shen Li, Shaohan Hu, Lu Su, Shuochao Yao, Huajie Shao, Hongwei Wang, and Tarek Abdelzaher. "Greendrive: A smartphone-based intelligent speed adaptation system with real-time traffic signal prediction." In Proceedings of the 8th International Conference on Cyber-Physical Systems, pp. 229-238. ACM, 2017.
57. Koukoumidis, Emmanouil, Li-Shiuan Peh, and Margaret Rose Martonosi. "SignalGuru: leveraging mobile phones for collaborative traffic signal schedule advisory." In Proceedings of the 9th international conference on Mobile systems, applications, and services, pp. 127-140. ACM, 2011.
58. Zhao, Yiran, Yang Zhang, Tuo Yu, Tianyuan Liu, Xinbing Wang, Xiaohua Tian, and Xue Liu. "CityDrive: A map-generating and speed-optimizing driving system." In INFOCOM, 2014 Proceedings IEEE, pp. 1986-1994. IEEE, 2014.
59. Alexander, W., X. Hong, and A. Hainen. "V2I Communication-Enabled Real-Time Intersection Traffic Signal Scheduling." In Proceedings of the SouthEast Conference, pp. 26-33. ACM, 2017.
60. Li, Xiaobing, Asad J. Khattak, and Airton G. Kohls. "Signal phase timing impact on traffic delay and queue length-a intersection case study." In Winter Simulation Conference (WSC), 2016, pp. 3722-3723. IEEE, 2016.
61. Treiber, Martin, Ansgar Hennecke, and Dirk Helbing. "Congested traffic states in empirical observations and microscopic simulations." Physical review E 62, no. 2 (2000): 1805.
62. Bachir, Abdelmalik, and Abderrahim Benslimane. "A multicast protocol in ad hoc networks inter-vehicle geocast." In Vehicular Technology Conference, 2003. VTC 2003-Spring. The 57th IEEE Semiannual, vol. 4, pp. 2456-2460. IEEE, 2003.
63. Van Aerde, M., B. Hellinga, M. Baker, and H. Rakha. "INTEGRATION: An overview of traffic simulation features." Transportation Research Records (1996).
64. Vuyyuru, Rama, Kentaro Oguchi, Clay Collier, and Ed Koch. "Automesh: Flexible simulation framework for vehicular communication." In Mobile and Ubiquitous Systems-Workshops, 2006. 3rd Annual International Conference on, pp. 1-6. IEEE, 2006.
65. Gelbal, Şükrü Yaren, Santhosh Tamilarasan, Mustafa Ridvan Cantaş, Levent Güvenç, and Bilin Aksun-Güvenç. "A connected and autonomous vehicle hardware-in-the-loop simulator for developing automated driving algorithms." In Systems, Man, and Cybernetics (SMC), 2017 IEEE International Conference on, pp. 3397-3402. IEEE, 2017.
66. Pariota, Luigi, Gennaro Nicola Bifulco, Gustav Markkula, and Richard Romano. "Validation of driving behaviour as a step towards the investigation of Connected and Automated Vehicles by means of driving simulators." In Models and Technologies for Intelligent Transportation Systems (MT-ITS), 2017 5th IEEE International Conference on, pp. 274-279. IEEE, 2017.
67. Zhao, Yunjie, Aditya Wagh, Kevin Hulme, Chunming Qiao, Adel W. Sadek, Hongli Xu, and Liusheng Huang. "Integrated traffic-driving-networking simulator: A unique R&D tool for connected vehicles." In Connected Vehicles and Expo (ICCVE), 2012 International Conference on, pp. 203-204. IEEE, 2012.
68. Kristensen, Terje, and Nnamdi Johnson Ezeora. "Simulation of intelligent traffic control for autonomous vehicles." In Information and Automation (ICIA), 2017 IEEE International Conference on, pp. 459-465. IEEE, 2017.
69. Timóteo, Ivo JPM, Miguel R. Araújo, Rosaldo JF Rossetti, and Eugenio C. Oliveira. "TraSMAPI: An API oriented towards Multi-Agent Systems real-time interaction with multiple Traffic Simulators." In Intelligent Transportation Systems (ITSC), 2010 13th International IEEE Conference on, pp. 1183-1188. IEEE, 2010.
70. Kim, Heejae, Jiyong Han, Seong-Hwan Kim, Jisoo Choi, Dongsik Yoon, Minsu Jeon, Eunjoo Yang et al. "IsV2C: an integrated road traffic-network-cloud simulator for V2C connected car services." In Services Computing (SCC), 2017 IEEE International Conference on, pp. 434-441. IEEE, 2017.